\documentclass[preprint, authoryear, 12pt]{elsarticle}
\usepackage{graphicx}
\usepackage{epstopdf}
\usepackage{amssymb}
\usepackage{amsmath}
\usepackage{hyperref}
\usepackage{natbib}
\usepackage{setspace}
\usepackage{color}
\usepackage{caption}
\usepackage{lineno}
\usepackage{makecell}
\usepackage{subfig}

\biboptions{authoryear, semicolon}

\journal{Icarus}

\begin{document}
\doublespacing
%\linenumbers

\begin{frontmatter}
\title{Ongoing Resurfacing of KBO Eris by Volatile Transport in Local, Collisional, Sublimation Atmosphere Regime} 

\author{Jason D. Hofgartner}
\cortext[tmp]{Corresponding author.}
\ead{Jason.D.Hofgartner@jpl.nasa.gov}
\author{Bonnie J. Buratti}
\address{Jet Propulsion Laboratory, California Institute of Technology, Pasadena, CA, USA}
\author{Paul O. Hayne}
\address{Department of Astrophysical and Planetary Sciences, University of Colorado, Boulder, CO, USA}
\author{Leslie A. Young}
\address{Southwest Research Institute, Boulder, CO, USA}

\begin{abstract}
Kuiper belt object (KBO) Eris is exceptionally bright with a greater visible geometric albedo than any other known KBO.  Its infrared reflectance spectrum is dominated by methane, which should form tholins that darken the surface on timescales much shorter than the age of the Solar System.  Thus one or more ongoing processes probably maintain its brightness.  Eris is predicted to have a primarily nitrogen atmosphere that is in vapor pressure equilibrium with nitrogen-ice and is collisional (not ballistic).  Eris's eccentric orbit is expected to result in two atmospheric regimes: (1) a period near perihelion when the atmosphere is global (analogous to the atmospheres of Mars, Triton, and Pluto) and (2) a period near aphelion when only a local atmosphere exists near the warmest region (analogous to the atmosphere of Io).  A numerical model developed to simulate Eris's thermal and volatile evolution in the local atmosphere regime is presented.  The model conserves energy, mass, and momentum while maintaining vapor pressure equilibrium.  It is adaptable to other local, collisional, sublimation atmospheres, which in addition to Io and Eris, may occur on several volatile-bearing KBOs.  The model was applied for a limiting case where Eris is fixed at aphelion and has an initial nitrogen-ice mass everywhere equal to the precipitable column of nitrogen in Pluto’s atmosphere during the New Horizons encounter (the resultant mass if the Pluto atmosphere collapsed uniformly onto the surface).  The model results indicate that (1) transport of nitrogen in the local, collisional, sublimation atmosphere regime is significant, (2) changes of Eris's albedo or color from nitrogen transport may be observable, and (3) uniform collapse of a global, nitrogen atmosphere likely cannot explain Eris's anomalous albedo in the present epoch.  Seasonal volatile transport remains a plausible hypothesis to explain Eris's anomalous albedo and geologic processes that renew Pluto's brightest surfaces, such as convection and glaciation, may also be operating on Eris.
\end{abstract}


\end{frontmatter}

\section{Motivation: The High Albedo of KBO Eris}

Eris is a large Kuiper belt object (KBO) with a radius that is 0.98 of Pluto's and a mass relative to Pluto of 1.27 \citep{Sicardy2011, Brown2007}.  It has an eccentric orbit, ranging from 38 AU to 98 AU from the Sun over a 557-year period.  Eris was at aphelion in 1977 and in 2018 is at a heliocentric distance of 96 AU.  Eris is also exceptionally bright with a visible geometric albedo of 0.96 \citep{Sicardy2011}, the highest geometric albedo of any known KBO.  Methane-ice dominates its infrared reflectance spectrum \citep{Brown2005}.  Shifts of the methane absorption features indicate bulk abundances of 90\% $N_2$ and 10\% $CH_4$ \citep{Tegler2010}.  Dysnomia is the only known satellite of Eris and has an orbital period of 16 days \citep{Brown2007}. 

\citet{Stern1988} predicted that a high albedo and methane-ice on Pluto (and by analogy other KBOs) implied recent or ongoing renewal of the surface because methane should form tholins that darken the surface on timescales much shorter than the age of the Solar System (see also \citet{Johnson1989}).  This prediction was confirmed in spectacular fashion by the New Horizons exploration of Pluto, which not only found that bright surfaces are also young surfaces but also that Pluto has extraordinarily bright surfaces that are also exceptionally young \citep{Stern2015, Buratti2017, Robbins2017}.  Temporal changes from ongoing surface renewal, however, were not observed on Pluto over the brief interval of the New Horizons flyby \citep{Hofgartner2018}.  Thus Eris's anomalously high albedo, along with the detection of methane-ice, strongly suggests that one or more ongoing processes maintain its brightness and resupply methane to its surface.

Atmospheric freeze-out (collapse onto the surface) as Eris recedes from its perihelion heliocentric distance of 38 AU to 98 AU at aphelion is commonly invoked to explain its high albedo (e.g., \citet{Brown2005, Sicardy2011}).  The $>$ 500-year period of Eris's orbit prohibits direct observational evaluation of this hypothesis but it can be investigated with a numerical thermal model.  We assess the hypothesis of atmospheric collapse of a Triton/Pluto-like nitrogen atmosphere for the anomalously high albedo of Eris using coupled thermal and transport models.  

In section~\ref{sec:AtmRegimes} we argue that Eris's eccentric orbit is expected to result in two atmospheric regimes: (1) a period near perihelion when the atmosphere is global and (2) a period near aphelion when only a local atmosphere exists near the warmest (approximately subsolar) region.  In both regimes the atmosphere is collisional (not ballistic), composed primarily of nitrogen, and is maintained by vapor pressure equilibrium (VPE) with solid-phase nitrogen (nitrogen-ice).  The global regime is analogous to the atmospheres of Mars, Triton, and Pluto and the local regime to that of Io.

Section~\ref{sec:Model} describes a model we developed to simulate thermal and volatile evolution in the local, collisional, sublimation atmosphere (LCSA)  regime.  The coupled thermal-transport model conserves energy, mass, and momentum while maintaining VPE.  The LCSA regime is unique to Io on large bodies closer to the Sun than KBOs and has received less study than other atmospheric regimes.  Several large KBOs are now known to have volatile deposits on their surfaces and expected to be in the local atmosphere regime for parts of their orbits \citep{Young2013}.  Thus, the local atmosphere regime may be common in the Solar System and will probably become a topic of increased study.  The model in section~\ref{sec:Model} is adaptable to other bodies with local atmospheres and is a possible starting point for future studies of their thermal and volatile evolutions.

In section~\ref{sec:Results} we demonstrate the implementation of the coupled thermal-transport model and use it to test the hypothesis of atmospheric collapse for the anomalously high albedo of Eris in a limiting case where Eris is held at aphelion.  This is followed by a discussion of the implications of the results of the model simulations and conclusions.

\section{Global and Local, Collisional, Sublimation Atmospheres}
\label{sec:AtmRegimes}

\begin{figure}[ht!]
\begin{center}
\includegraphics[width=0.49\textwidth]{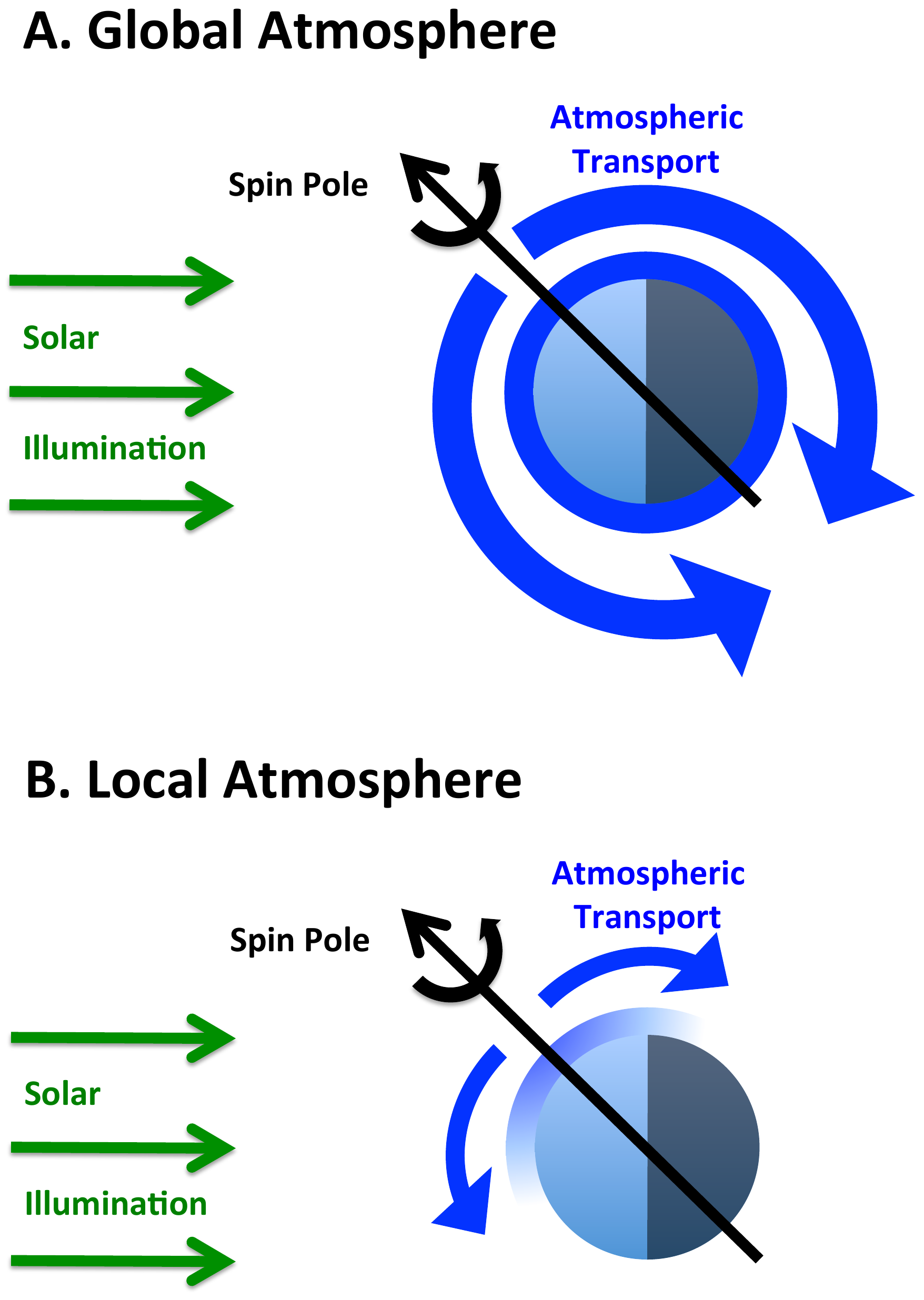}
\end{center}
\caption{Conceptual diagrams of global and local, collisional, sublimation atmospheres.  (A.) In the global atmosphere regime, an atmosphere with approximately uniform surface pressure (isobaric) covers the globe.  Illuminated volatile-ice generally undergoes sublimation while condensation generally occurs in dark regions.  The atmosphere transports the volatile from the sublimating to the condensing regions so that the pressure remains isobaric.  (B.) In the local atmosphere regime, the condensation and/or sublimation rates exceed the rate of atmospheric transport and atmospheric transport cannot maintain a globally uniform surface pressure.  The temperature of the volatile-ice is not held isothermal by VPE with an isobaric atmosphere.  A local atmosphere occurs over the warmest volatile-ice covered region but the atmospheric pressure decreases over colder volatile-ice away from this region.  The pressure gradients generate winds that transport the volatile away from the warmest region.  Due to the significant decrease in vapor pressure away from the warmest region, the transport in the local atmosphere regime may not include the whole globe.}
\label{AtmDiagrams}
\end{figure}

We refer to an atmosphere where the primary constituent is in VPE with solid-phase surface deposits (volatile-ice) as a sublimation atmosphere.  A global sublimation atmosphere has approximately uniform surface pressure everywhere around the globe (isobaric); due to VPE, the volatile-ice has an approximately uniform temperature around the globe (isothermal; Fig.~\ref{AtmDiagrams}; \citet{Trafton1983}).  Condensation occurs where energy balance requires latent heat to be added (generally at the winter pole) and sublimation where latent heat is subtracted (summer pole).  If the sublimating mass is greater than the condensing mass, the pressure of the atmosphere increases everywhere and vice versa.  Mars, Triton, and Pluto have global sublimation atmospheres.  Note that these are different from Earth's atmosphere because its primary constituent is always in the vapor phase.  Water condenses/sublimates toward VPE but it is a minor constituent that does not hold the pressure isobaric and the condensed phases isothermal.  

If temperature/pressure changes in one region are much faster than the time required to communicate those changes over the globe, VPE cannot sustain a global atmosphere.  We approximate the timescale of local pressure changes as the timescale for atmospheric freeze-out (radiative collapse), which is estimated as the ratio of the latent heat that must be emitted to condense a column of atmosphere to the radiative flux.  The timescale for communicating pressure changes is approximately the pole-to-pole distance divided by the sound speed.  We define $\alpha$ as the ratio of these two timescales\begin{equation} \alpha \equiv \frac{t_{freeze-out}}{t_{pole-to-pole}} \approx \frac{\Sigma L/(\epsilon\sigma T^4)}{\pi R/c_s} \approx \frac{PL/(g\epsilon\sigma T^4)}{\pi R/\sqrt{N_A kT/\mu}} \end{equation} where $\Sigma$ is mass of atmosphere per unit area, $L$ is latent heat of sublimation/condensation, $\epsilon$ is surface emissivity, $\sigma$ is the Stefan-Boltzmann constant, $T$ is volatile-ice surface temperature, $R$ is planetary radius, $c_s$ is speed of sound, $P$ is vapor pressure, $g$ is gravitational acceleration, $N_A$ is the Avogadro constant, $k$ is the Boltzmann constant, and $\mu$ is molecular mass.  If $\alpha >> 1$ then a sublimation atmosphere is global.  If $\alpha << 1$, atmospheric transport cannot keep up with condensation in dark regions and an atmosphere will exist only locally over sublimating regions (Fig.~\ref{AtmDiagrams}).  This atmosphere will have pressure gradients that will result in horizontal transport of the vapor away from the warmest regions.  We refer to this regime as a local sublimation atmosphere.  

\citet{Trafton1983} argued that a sublimation atmosphere would be in the global regime if the wind speed required to hold the volatile-ice isothermal is much less than the speed of sound.  Assuming that the whole globe is covered by isothermal volatile-ice, they also showed that the steady-state wind speed $v$, is \begin{equation} v = \frac{R\epsilon\sigma T^4}{\Sigma L}. \end{equation}  For this special case, $\alpha = c_s/(\pi v)$ and the condition that $\alpha >> 1$ for a sublimation atmosphere to be global is equivalent within a factor of $\pi$ to the condition that the wind speed be much less than the speed of sound ($v << c_s$).  

Table~\ref{tbl:AlphaBeta} gives $\alpha$ for several planetary bodies known to have sublimation atmospheres.  For Mars, Triton, and Pluto, we find $\alpha \approx$ 20, 90, and 200 respectively, consistent with their global atmospheres, and for Io, $\alpha \approx 3 \times 10^{-5}$, consistent with its local atmosphere.  We note that $\alpha$ for Triton and Pluto during the Voyager 2 and New Horizons encounters is greater than the $\alpha$ for Mars, suggesting that at those times their atmospheres were more stable against collapse than the atmosphere of Mars.  When Pluto is at aphelion, however, $\alpha \approx 1$ (assuming the same emissivity and albedo to calculate the planetary equilibrium temperature) and thus its atmosphere is less stable against collapse.  The atmosphere is less stable against collapse but since $\alpha \approx 1$ it does not provide a good indication of whether or not the atmosphere does collapse; the parameter is only useful for distinguishing between global and local atmospheres when $\alpha >> 1$ or $\alpha << 1$.  Research focused on the stability of Pluto's atmosphere suggests that a high thermal inertia prevents collapse of its global sublimation atmosphere (e.g.,\citet{Young2013b, Olkin2015}). 

\begin{table}[!htbp]
\begin{center}
\caption{Parameters for estimating whether a sublimation atmosphere is global or local ($\alpha$) and collisional or ballistic ($\beta$).  Since these are estimates, the parameters are only quoted to one significant digit.  For Eris, the global-average planetary equilibrium temperature (for surface bolometric Bond albedo of $A$ and assuming an emissivity of unity for simplicity) and the corresponding vapor pressure of the volatile ($N_2$ or $CH_4$; from \citet{Fray2009}) were used.}

\begin{tabular}{|l|c|c|}
\hline
 & $\alpha$ & $\beta$ \\
\hline
Mars ($CO_2$) & 20 & $1 \times 10^{9}$ \\
Io ($SO_2$) & $3 \times 10^{-5}$ & 200 \\
Triton ($N_2$) & 90 & $2 \times 10^{7}$ \\
Pluto ($N_2$) & 200 & $3 \times 10^{7}$ \\
Eris ($N_2$, perihelion, A=0.9) & 0.008 & 200 \\
Eris ($N_2$, aphelion, A=0.9) & $1 \times 10^{-11}$ & $8 \times 10^{-8}$ \\
Eris ($N_2$, perihelion, A=0.5) & 200 & $3 \times 10^{7}$ \\
Eris ($N_2$, aphelion, A=0.5) & $7 \times 10^{-4}$ & 20 \\
Eris ($N_2$, perihelion, A=0.1) & 2000 & $7 \times 10^{8}$ \\
Eris ($N_2$, aphelion, A=0.1) & 0.09 &  3000 \\
Eris ($CH_4$, perihelion, A=0.9) & $3 \times 10^{-8}$ & 0.001 \\
Eris ($CH_4$, aphelion, A=0.9) & $2 \times 10^{-19}$ & $9 \times 10^{-16}$ \\
Eris ($CH_4$, perihelion, A=0.5) & 0.04 & 5000 \\
Eris ($CH_4$, aphelion, A=0.5) & $2 \times 10^{-9}$ & $4 \times 10^{-5}$ \\
Eris ($CH_4$, perihelion, A=0.1) & 2 & $4 \times 10^{5}$\\
Eris ($CH_4$, aphelion, A=0.1) & $9 \times 10^{-7}$ & 0.03 \\
\hline
\end{tabular}
\label{tbl:AlphaBeta}
\end{center}
\end{table}

Methane dominates the reflectance spectrum of Eris \citep{Brown2005}; however, shifts of the methane absorption features indicate bulk abundances of $\approx$ 10\% $CH_4$ and 90\% $N_2$, similar to the bulk abundances on Pluto \citep{Tegler2010}.  Thus nitrogen is likely the primary constituent of any atmosphere on Eris, as it is on Triton and Pluto, and dominates the latent heat of sublimation.  The visible geometric albedo of Eris is known but not its bolometric Bond albedo; these parameters are related but not uniquely.  The Bond albedo is equal to the geometric albedo multiplied by the phase integral and if Eris's phase integral is similar to that of the brightest Saturnian satellites, then $A \approx 0.5-0.7$ \citep{Sicardy2011}.  Since the present bolometric Bond albedo is not known and the albedo could change in time (from change of the illuminated hemisphere or surface evolution), we consider a range of bolometric Bond albedos.  For nitrogen on Eris, at aphelion we find $\alpha \approx 0.1 - 10^{-11}$ for bolometric Bond albedos of $0.1 - 0.9$ respectively  and at perihelion $\alpha \approx 2000 - 0.008$.  Eris is therefore predicted to have a local atmosphere at aphelion and probably transition to a global atmosphere as it approaches perihelion, depending on its bolometric Bond albedo.  If methane is taken as the dominant constituent of any atmosphere on Eris, we find that Eris is likely always in the local atmosphere regime, except possibly for very dark albedos right at perihelion.  This difference is due to the lower volatility of methane compared to nitrogen (e.g., \citet{Fray2009}) and demonstrates that if nitrogen is present in sufficient quantity on Eris, it will be the main constituent of the atmosphere.

We refer to an atmosphere as ballistic if vapor particles have no interaction with each other and follow ballistic trajectories; the converse is a collisional atmosphere.  To check if an atmosphere is collisional, a useful metric is the ratio of the pressure scale height to the mean free path.  We define $\beta$ to be this ratio \begin{equation} \beta \equiv \frac{H}{l} \approx \frac{N_A kT / (\mu g)}{kT / (\sqrt{2}\pi D^2 P)} \end{equation}  where $H$ is pressure scale height, $l$ is mean free path, and $D$ is kinetic diameter (diameter of effective scattering cross section). If $\beta >> 1$ then the atmosphere is collisional and if $\beta << 1$ then the atmosphere is ballistic.  Note that $\beta$ is the inverse of the Knudsen number.  

For Mars, Io, Triton, and Pluto we find, $\beta \approx 1 \times 10^9, 200, 2 \times 10^7,$ and $3 \times 10^7$ respectively, consistent with their collisional atmospheres.  For nitrogen on Eris, $\beta \approx 10^{-7} - 10^{9}$ depending on albedo and solar distance (Table~\ref{tbl:AlphaBeta}).  We note that $\beta$ is $< 1$ for Eris only for the very coldest planetary equilibrium temperatures.  Since the temperature of the subsolar point is greater than the planetary equilibrium temperature, even for high albedos and at aphelion Eris likely does have a collisional atmosphere at the subsolar point.  For methane on Eris, $\beta << 1$ except for the highest planetary equilibrium temperatures.

We therefore predict that Eris has a nitrogen, LCSA for part if not all of its orbit and possibly a nitrogen, global, collisional, sublimation atmosphere when it is closest to the Sun.

\section{Coupled Thermal-Transport Numerical Model of Local, Collisional, Sublimation Atmospheres}
\label{sec:Model}

\subsection{Numerical Thermal Model of Global, Collisional, Sublimation Atmospheres}

Volatile transport (VT) in the global, collisional, sublimation atmosphere regime was modeled with a numerical thermal model that conserves energy and mass while maintaining VPE.  \citet{Leighton1966} developed this model to study the behavior of carbon dioxide on Mars and correctly predicted (1) that the Martian seasonal polar caps are composed primarily of carbon dioxide and (2) that the total pressure of the Martian atmosphere changes substantially over an annual cycle from exchange of carbon dioxide between the polar caps and atmosphere.  The model was adapted to study the seasonal nitrogen cycles on Triton and Pluto and demonstrated that their atmospheres are strongly connected to nitrogen-ice on their surfaces by VPE, analogous to carbon dioxide on Mars (e.g.,\citet{Hansen1992, Hansen1996}).  This type of thermal model was used extensively to study thermal and volatile behavior on Mars, Triton, and Pluto, and the code for a refined version of the model that is computationally efficient for a broad phase space of initial conditions is publicly available \citep{Young2017}.  

In the global atmosphere regime, atmospheric transport is able to keep pace with pressure changes from sublimation/condensation and maintains approximately uniform pressure over the globe.  The models described above assume a uniform pressure (and by VPE, a uniform volatile-ice temperature), which implicitly incorporates transfer of energy and mass, so these models do not explicitly track the atmospheric transport.  These models also include subsurface conduction, but subsurface conduction can be ignored if the surface is covered by isothermal volatile-ice.  For this simplified case, these models have three variables: volatile-ice temperature $T$, vapor pressure $P$, and volatile-ice mass per unit area $m$, all of which vary with time but only $m$ varies with position.  The variables are determined by three equations: conservation of energy and mass and VPE; only the equation for conservation of energy varies with position (solved locally at each position).  The assumption of uniform pressure is not valid in the local atmosphere regime and atmospheric transport of mass and energy must be explicitly modeled.

\subsection{Meteorological Model of Local, Collisional, Sublimation Atmospheres}

VT in the LCSA regime was modeled with a vertically integrated meteorological model that conserves energy, mass, and momentum.  \citet{Ingersoll1985} developed this model to study the behavior of sulfur dioxide on Io.  Their model has three variables: $T$, $P$, and wind speed $v$, all of which vary with position, and three equations, all of which also vary with position.  The model prescribed the volatile-ice temperature distribution, did not include latent heat, and assumed symmetry about the subsolar/antisolar axis.  As a result, both the thermal and temporal behavior were not explicitly modeled.  While all of these assumptions were shown to be reasonably appropriate for Io, none are valid for a general LCSA.
  
\subsection{Coupled Thermal-Transport Numerical Model of Local, Collisional, Sublimation Atmospheres}

\begin{figure}[ht!]
\begin{center}
\includegraphics[width=\textwidth]{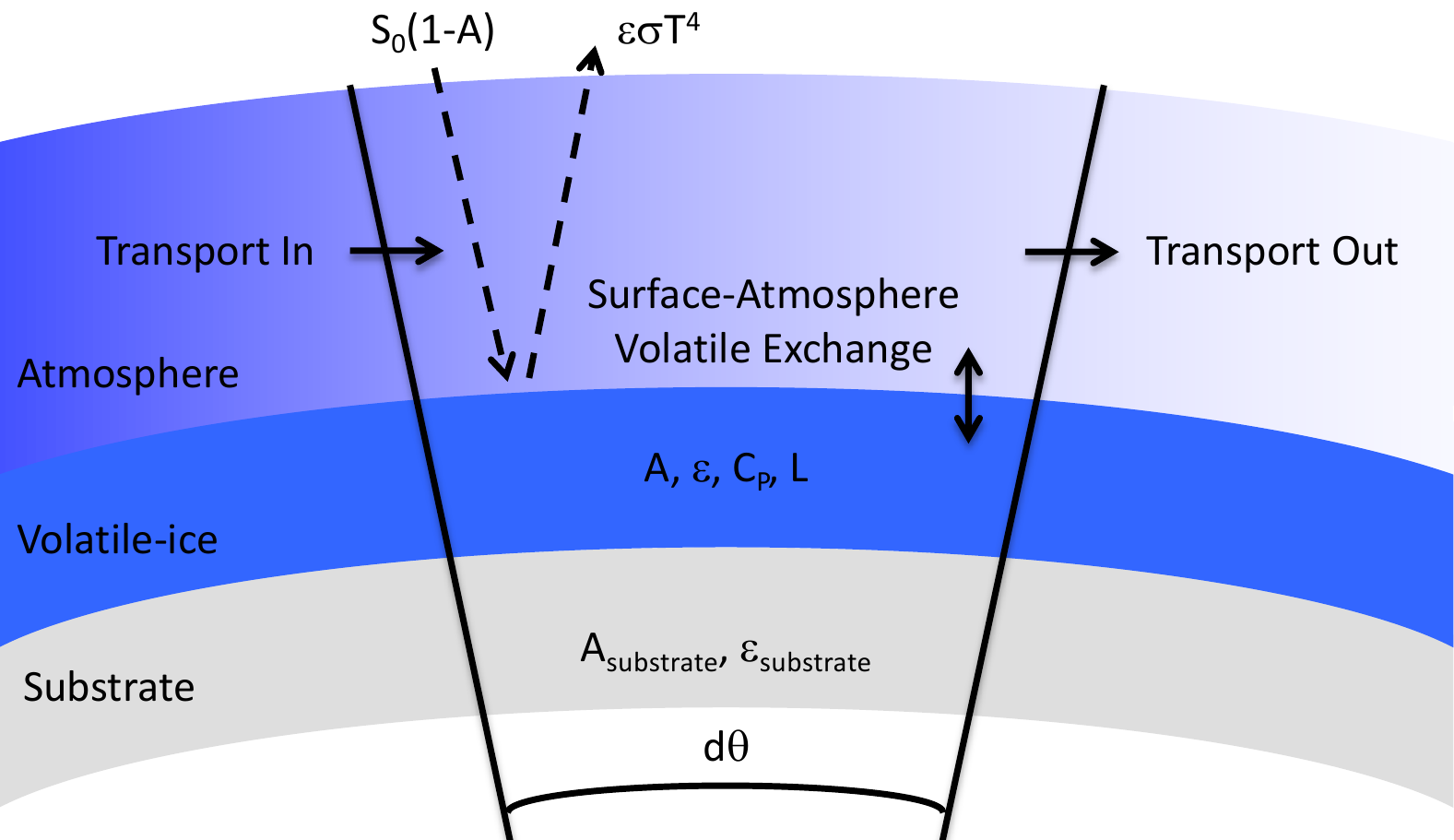}
\end{center}
\caption{Conceptual diagram of coupled thermal-transport numerical model for the local, collisional, sublimation atmosphere regime.  The diagram does not include every component of the model; see the text for the complete model.  The radiation terms in the equation for energy conservation are depicted by dashed arrows and accompanying expressions.  The heat capacity and latent heat terms are noted by constants in the volatile-ice layer.  Conservation of mass is depicted by the three solid arrows and accompanying labels.  A gradient in atmospheric pressure is depicted by a gradient in the color of the atmosphere.  The substrate in the diagram is covered by a uniform thickness volatile-ice layer, however, the model includes variable volatile-ice thickness.}
\label{MassBalance}
\end{figure}

We combine elements of the two types of models discussed above to model the LCSA regime with a coupled thermal-transport numerical model.  As in the global thermal model, energy and mass are conserved while maintaining VPE, but for the local regime all of these equations vary with position and are solved at each location individually rather than averaged globally.  As in the local meteorological model, the atmospheric transport of volatile material due to pressure gradients is tracked explicitly via the conservation of momentum.  The transport of mass and energy from this atmospheric transport is added to their respective conservation equations.  The coupled thermal-transport model conserves energy, mass, and momentum while maintaining VPE.  Fig.~\ref{MassBalance} depicts some, but not all, of the components of this model.  The model is adaptable to any planetary body with a LCSA.

\textbf{Conservation of Energy:}  The difference between absorbed and emitted radiation is balanced by changes of temperature and phase, \begin{equation} S_0(1-A) - \epsilon\sigma T^4 = mC_{P}\frac{dT}{dt} - L\frac{dm}{dt}, \end{equation} where $S_0$ is incident solar flux, $A$ is surface bolometric Bond albedo, $\epsilon$ is surface emissivity, $\sigma$ is the Stefan-Boltzmann constant, $T$ is surface temperature, $m$ is mass of solid-phase volatile deposit (volatile-ice) per unit area, $C_P$ is volatile-ice heat capacity at constant pressure, $t$ is time, and $L$ is latent heat of volatile-ice sublimation. $S_0$, $T$, and $m$ vary with position and time where $T$ and $m$ are variables to be solved and $S_0$ is known (see for example, \citet{Sanchez2011} for the expression for $S_0$ as a function of position and time).  

The heat capacity of solid-phase nitrogen for relevant temperatures is given in \citet{Scott1976} and we use those values in this work.  As in \citet{Hansen1996}, the model tracks whether the nitrogen-ice is in its $\alpha$ (cubic crystal structure) or $\beta$ (hexagonal crystal structure) phase.  It uses the values provided therein for the latent heat of sublimation of each phase and the latent heat of the $\alpha-\beta$ phase change. 

Internal energy sources and conduction between the surface and subsurface could be incorporated into the equation for conservation of energy as done in other thermal models.  However, for surfaces covered by volatile-ice in VPE with the atmosphere, subsurface conduction is a minor term in the equation and can be ignored.  The present-day radiogenic heat flux to Pluto's surface is estimated to be $\approx 0.003$ W/m$^2$ (e.g., \citet{McKinnon2016}).  Eris should not have a drastically different heat flux.  Assuming Pluto and Eris have two component interiors with densities of 1000 kg/m$^3$ (water-ice) and 3500 kg/m$^3$ (rock), that all of their radiogenic energy comes from their rocky material, and that the radiogenic power per unit mass of rock is the same for both bodies, the present-day radiogenic heat flux to Eris's surface is $\approx 0.005$ W/m$^2$.  The solar flux at 98 AU is $\approx 0.14$ W/m$^2$.  Thus even at Eris's great aphelion distance, solar energy significantly exceeds the expected internal energy and therefore neglecting the internal contribution should only modestly affect the results.  Similarly, the heat capacity of the atmosphere is another term that could be included in the equation for conservation of energy.  But, for the low atmospheric pressures on Eris, the magnitude of this term is always $<$ 2\% the magnitude of the latent heat term, which is less than the uncertainty in the latent heat of sublimation.  We have decided not to include internal energy, subsurface conduction, and atmospheric heat capacity in this first application of the model because they are minor terms in the equation for conservation of energy and would introduce additional assumptions and/or variables.

\textbf{Conservation of Mass:}  The volatile-ice deposits and atmosphere exchange mass through condensation/sublimation while conserving their total mass.  Since the mass of atmosphere per unit area is $\approx P/g$ where $P$ is vapor pressure and $g$ is gravitational acceleration, \begin{equation} \frac{dm}{dt} \approx -\frac{1}{g}\frac{dP_{exchange}}{dt}, \end{equation} where $dP_{exchange} > 0$ corresponds to volatile-ice sublimation and $dP_{exchange} < 0$ to condensation.  The total change of pressure at each position is the sum of the contributions from exchange with the volatile-ice and atmospheric transport, \begin{equation} dP = dP_{exchange} + dP_{transport} \approx -g(dm) + \frac{1}{R\sin\theta}\frac{\partial}{\partial\theta}(v_\theta P\sin\theta)dt, \end{equation} where $R$ is planetary radius, $\theta$ is polar angle from the location with the highest pressure, and $v_\theta$ is the $\theta$ vector component of the wind speed ($v_\theta > 0$).  The second term after the last equality is the vertically integrated divergence of the meridional mass flux, as in \citet{Ingersoll1985}.  We have ignored the zonal mass flux since we're interested in seasonal meridional transport and the zonal component averages to zero over a diurnal cycle.  The pressure of the atmosphere is determined by VPE so \begin{equation} dP = \left(\frac{dP}{dT}\right)_{VPE}dT. \end{equation}  From the above two equations, the equation for conservation of mass is \begin{equation} \frac{dm}{dt} = -\frac{1}{g}\left(\left(\frac{dP}{dT}\right)_{VPE}\frac{dT}{dt} - \frac{1}{R\sin\theta}\frac{\partial}{\partial\theta}(v_\theta P\sin\theta) \right). \label{MassConservation} \end{equation}    

\textbf{Conservation of Momentum:}  Newton's second law for a fluid is, \begin{equation} \rho\frac{D\vec{v}}{Dt} = \rho\left(\frac{\partial v}{\partial t} + \vec{v} \cdot \nabla \vec{v}\right) = -\nabla P, \end{equation} where $\rho$ is the vapor density which we determine from the equation of state for an ideal gas.  If we again vertically integrate the meridional component and ignore the zonal component this equation simplifies to, \begin{equation} \rho v_\theta\frac{\partial v_\theta}{\partial \theta} = -\frac{\partial P}{\partial \theta}. \end{equation} 

\textbf{Vapor Pressure Equilibrium:} \citet{Fray2009} provide the pressure-temperature relation of VPE for $N_2$ (among many other volatiles of astrophysical interest) for the relevant temperatures and we use their relation.  

Thus the model has four variables: $T$, $P$, $m$, and $v_\theta$, all of which vary with position and time, and four equations: conservation of energy, mass, and momentum and VPE, all of which vary with position.

\section{Volatile Transport on Eris in the Local, Collisional, Sublimation Atmosphere Regime}
\label{sec:Results}

In the first application of the coupled thermal-transport numerical model we test the hypothesis of atmospheric freeze-out (radiative collapse) for the anomalously high albedo of Eris (e.g., \citet{Brown2005, Sicardy2011}).  Since Eris's perihelion distance is similar (within several AU) to that of Pluto's heliocentric distance during the New Horizons encounter, the volatile-ice deposit from atmospheric collapse can be crudely estimated by assuming Eris had a Pluto-like atmosphere when it was near perihelion and it subsequently condensed uniformly over the surface.  We use an initial volatile-ice mass equal to the precipitable nitrogen in Pluto's atmosphere as measured by New Horizons, $\approx 1.9$ kg/m$^2$ (corresponds to a nitrogen-ice layer $\approx 2$ mm thick; \citet{Young2018}).  In the simulations, Eris is held at its aphelion distance, where the incident solar flux is a minimum, as a limiting case.  If, even in this extreme limit, VT is not negligible, then VT must be significant throughout the orbit of Eris.  Similarly, we assume a friction force between the atmosphere and surface that reduces the wind speeds by a factor of 10.  Greater wind speeds result in greater VT.  The Bond albedo is equal to the geometric albedo multiplied by the phase integral and if a phase integral similar to the brightest Saturnian satellites is assumed then $A \approx 0.5-0.7$ \citep{Sicardy2011}.  For consistency with other studies of thermal and volatile evolution of KBOs, an emissivity of 0.9 is assumed (e.g., \citet{Stansberry2008}).  The rotation period of Eris was considered in several projects, however, there is significant disagreement, ranging from approximately 14 hours to 16 days (e.g. \citet{Carraro2006, Sheppard2007, Maris2008, Duffard2008, Roe2008, Rabinowitz2014}).  We consider this parameter to be presently unknown and for simplicity choose a rotation period equal to the orbital period of its satellite Dysnomia, $\approx 16$ Earth days \citep{Rabinowitz2014}.  Similarly, Eris's rotational pole is unknown so we assume it is equal to the pole of Dysnomia's orbit which corresponds to an obliquity of $78^\circ$ and a subsolar latitude of $\approx 40^\circ$ in the current epoch (recall that Eris is currently near aphelion; \citet{Brown2007}).  There is uncertainty in all of these initial conditions and parameters but the primary purposes of this first application of the coupled thermal-transport model are: (1) to demonstrate the applicability of the model and (2) determine whether VT in the local atmosphere regime on Eris is significant or negligible.  Careful consideration of the VT as these parameters and initial conditions are varied is left to a follow-on study. 

\begin{figure}[ht!]
  \centering
  \subfloat[Volatile-ice Temperature]{\includegraphics[width=0.5\textwidth]{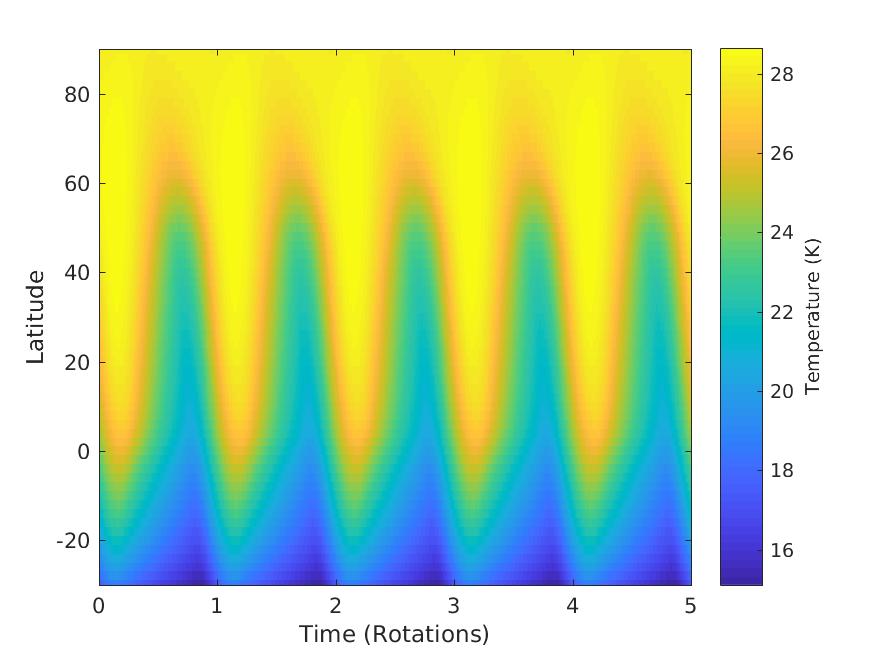}}                
  \subfloat[Vapor Pressure]{\includegraphics[width=0.5\textwidth]{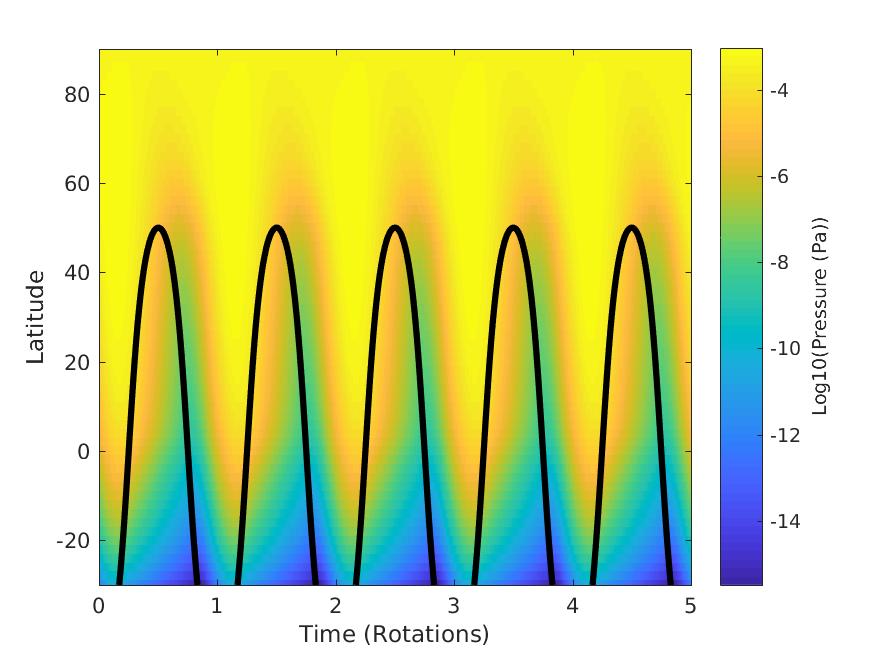}}   
\newline             
  \subfloat[Meridional Wind Speed]{\includegraphics[width=0.5\textwidth]{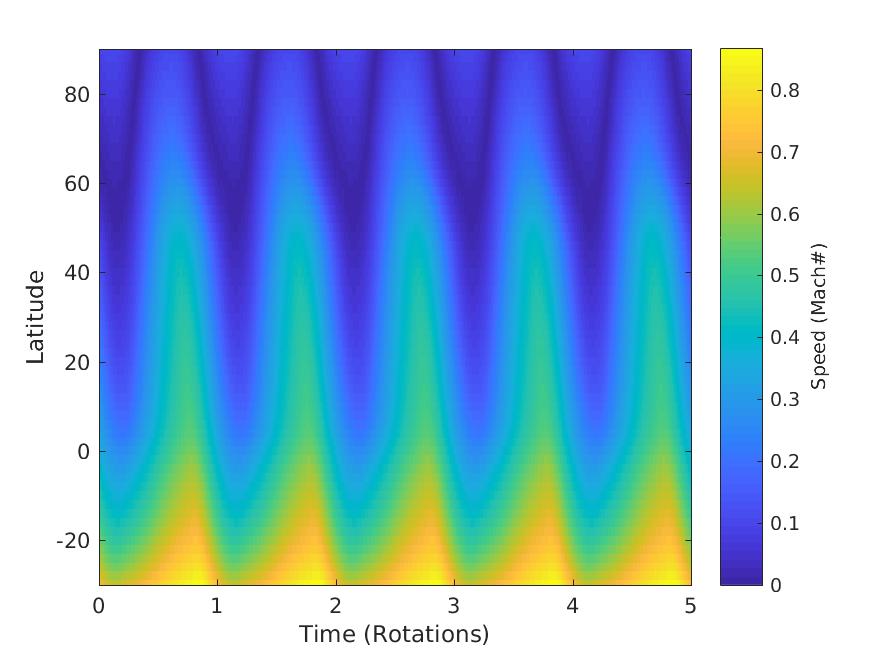}}                
  \subfloat[Volatile-ice Mass Per Unit Area]{\includegraphics[width=0.5\textwidth]{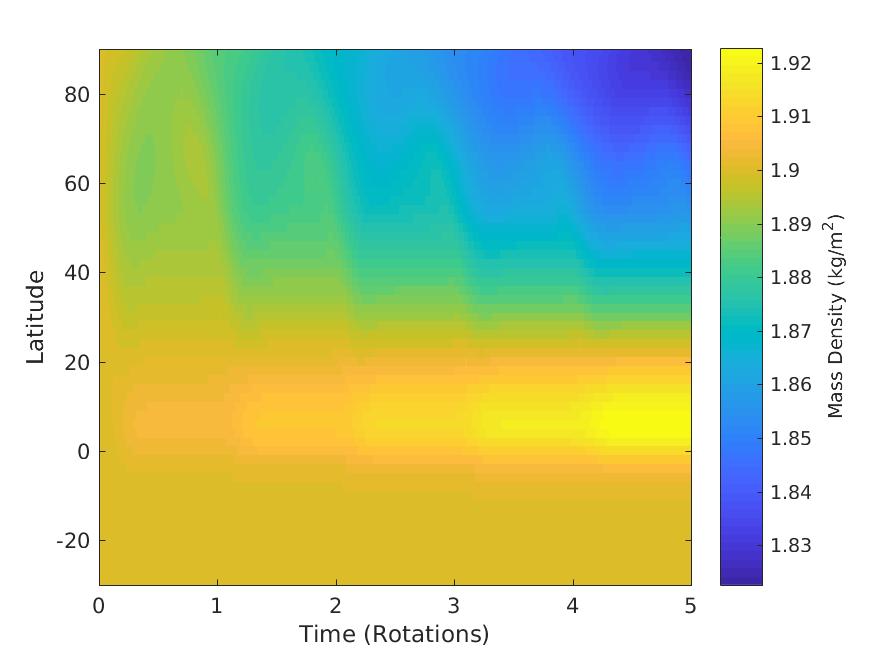}}                
\caption{Coupled thermal-transport numerical model results for Eris.  The surface temperature of the volatile-ice (a), atmospheric pressure at the surface (b), meridional wind speed (c), and volatile-ice mass per unit area (d) are shown as functions of latitude and time (due to zonal symmetry, longitudinal variability is equivalent to temporal variability over one rotation).  For this example simulation, Eris is held at its aphelion distance ($\approx$ 98 AU), has a rotation period equal to the orbital period of its satellite Dysnomia ($\approx$ 16 Earth days, \citet{Rabinowitz2014}), a subsolar latitude of $40^\circ$ (rotation pole equal to Dysnomia-orbit pole; \citet{Brown2007}), and the volatile-ice has a bolometric Bond albedo and emissivity of 0.6 and 0.9 respectively \citep{Sicardy2011, Stansberry2008}.  Friction is assumed to decrease the meridional wind speed of the atmosphere to 10\% of the frictionless speed.  The initial volatile-ice mass is everywhere equal to the precipitable column of nitrogen in Pluto's atmosphere during the New Horizons encounter (the resultant mass if the Pluto atmosphere collapsed uniformly over the globe).  This example demonstrates that the model results are reasonable.}
\label{fig:ModelOut}
\end{figure}

The four model outputs: volatile-ice temperature $T$, vapor pressure $P$, meridional wind speed $v_\theta$, and volatile-ice mass per unit area $m$, are shown in figure~\ref{fig:ModelOut}.  The output nitrogen-ice temperatures are reasonable, there is a diurnal oscillation, the maximum is near the subsolar latitude shortly after local noon, and the values are comparable to the planetary equilibrium temperature ($\approx$ 23 K).  There is, however, a diurnal oscillation at the pole that is not realistic for the case of zonal symmetry.  This oscillation is a result of ignoring zonal transport in the model.  If all longitudes were included in the model, the contribution to the transport term at the pole in equation~\ref{MassConservation} from each longitude would oscillate with time of day but their sum would be a constant.  Since not all longitudes are modeled, the transport term at the pole is not a constant but oscillates with time of day resulting in an oscillation of the volatile-ice temperature.  Since we're primarily interested in the net transport away from the pole and the oscillation is small (nearly imperceptible in figure~\ref{fig:ModelOut}) this unphysical oscillation does not affect the results and conclusions. 

The vapor pressures in figure~\ref{fig:ModelOut} are similarly reasonable.  The pressure of the atmosphere at the surface is constrained by a stellar occultation to be $< 10^{-4}$ Pa (1$\sigma$ confidence level) over the limb of the disk but could be greater away from the limb \citep{Sicardy2011}.  The terminator, approximately the limb (since the subsolar and sub-Earth point on Eris when it is at aphelion are separated by $< 1^\circ$), is the black line in the figure.  The predicted pressures at the limb are consistent with the upper limit, ranging from $\approx 10^{-4}$ Pa, equal to the upper limit, to many orders of magnitude smaller depending on latitude.  The meridional wind speeds range from zero at the location of the temperature/pressure maximum (an imposed boundary condition as in \citet{Ingersoll1985}) to approximately half the sound speed.  In their study of Io's LCSA, \citet{Ingersoll1985} predicted wind speeds a few times the sound speed and concluded that supersonic speeds were a general feature of the transport.  Since the simulation in figure~\ref{fig:ModelOut} assumed that friction decreased the meridional wind speed of the atmosphere to 10\% of the frictionless speed, our results are similar to the predictions for Io.  The volatile-ice mass oscillates due to exchange of nitrogen between the volatile-ice and atmosphere (more of the nitrogen is in the vapor phase on the day side, when temperatures are higher, and vice versa).  There is also a net transport of nitrogen-ice from latitudes near the summer pole toward latitudes near the equator.  The net transport does not extend all the way to the winter pole (the coldest region), however, because the pressures become too small to transport appreciable quantities of the volatile.  Thus all four of the output variables are well-behaved and reasonable.

\begin{figure}[ht!]
  \centering
  \subfloat[$A=0.6, l_s = 40^\circ$]{\includegraphics[width=0.5\textwidth]{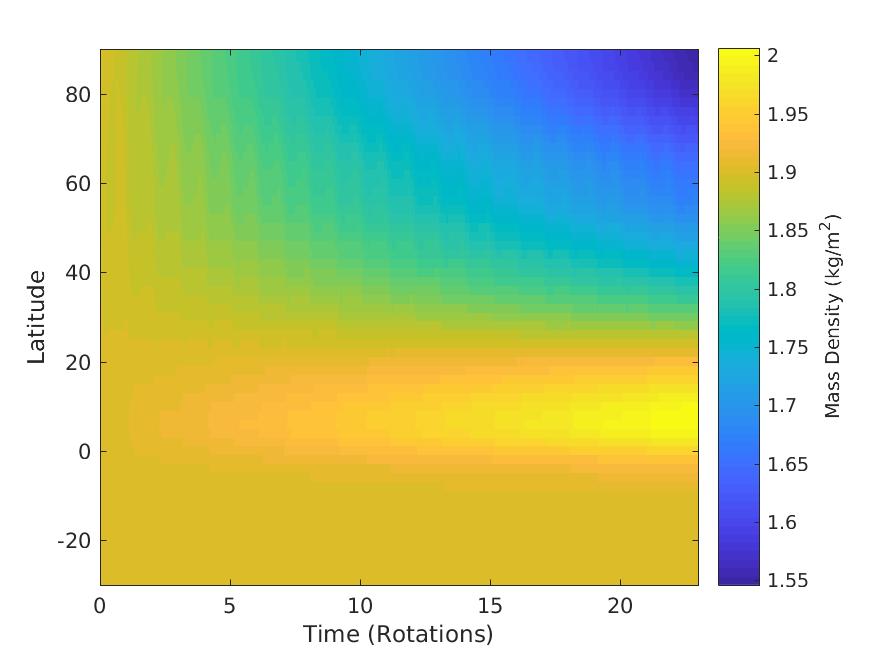}}                
  \subfloat[$A=0.5, l_s = 40^\circ$]{\includegraphics[width=0.5\textwidth]{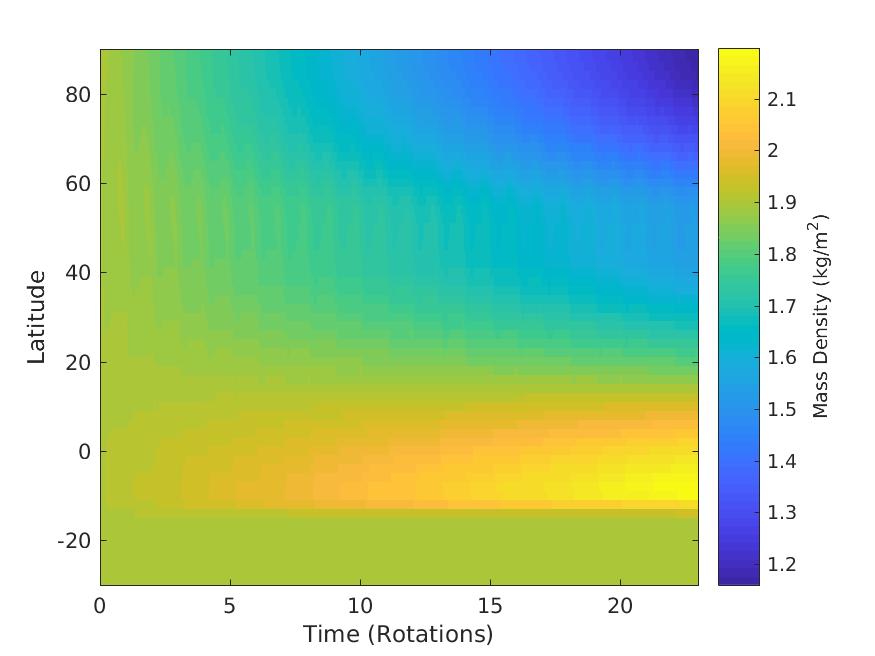}}                
\newline             
  \subfloat[$A=0.6, l_s = 50^\circ$]{\includegraphics[width=0.5\textwidth]{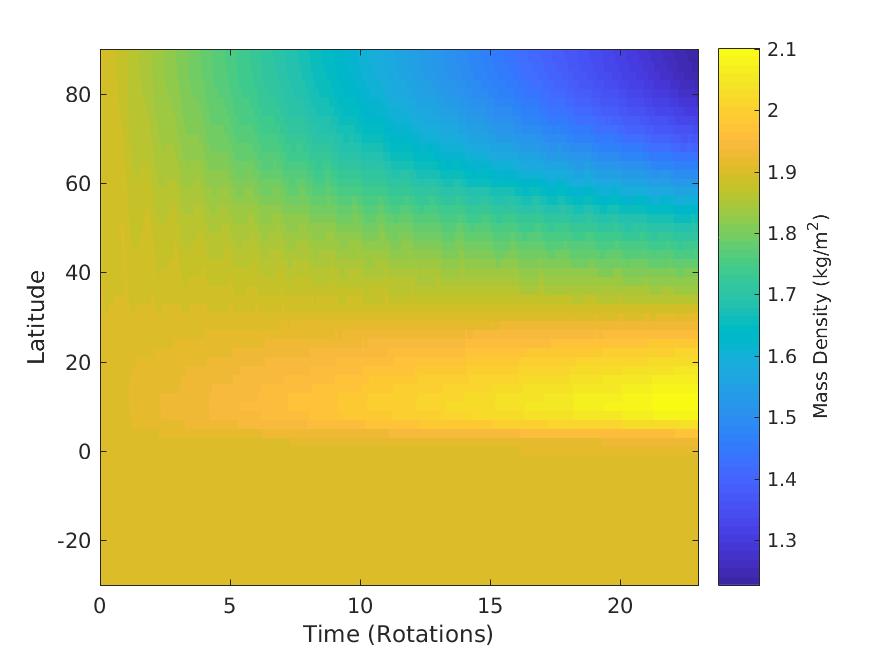}}                
\caption{Panel (a) shows the nitrogen-ice mass per unit area for 23 rotation periods (approximately one Earth-year) for the same initial conditions as in figure~\ref{fig:ModelOut}.  Significant volatile transport occurs over this period indicating that volatile transport is not a negligible process even when Eris is in the local atmosphere regime and at aphelion.  Panels (b) and (c) show the nitrogen-ice mass per unit area for the same time period with one modification to the model parameters that results in slight changes in the temperature distribution (the maximum temperature increases by $\approx$ 1 K).  The change in mass transport in the panels, however, is significantly greater (the mass-loss rate from the pole approximately doubles).  These panels demonstrate that volatile transport can depend sensitively on the model parameters and increasing temperature increases the role of atmospheric transport.}
\label{fig:Mass}
\end{figure}

Figure~\ref{fig:Mass} (a) shows the nitrogen-ice mass per unit area for 23 rotation periods (approximately one Earth-year) for the same initial conditions as in figure~\ref{fig:ModelOut}.  After $\approx$ 1 year the pole has lost an appreciable amount ($\approx$ 20\% or 400 $\mu$m thick layer) of nitrogen-ice to the equator.  In this simulation the pole of Eris is free of nitrogen-ice after less than 6 Earth-years.  Thus for the limiting case described above and assuming reasonable parameters, VT on Eris can significantly alter the surface on decadal timescales.  Even in the local atmosphere regime, at the very cold temperatures of an almost 100 AU solar distance, VT could be an important process on Eris.  The importance of VT for the surface evolution increases at closer heliocentric distances, where greater insolation increases volatile-ice temperature and the corresponding vapor pressure which results in greater atmospheric transport.  

The VT rate is substantially higher for some modest changes to the model parameters.  Panels (b) and (c) of figure~\ref{fig:Mass} show the nitrogen-ice mass per unit area when the bolometric Bond albedo of the nitrogen-ice is decreased from 0.6 to 0.5 or the subsolar latitude is increased from $40^\circ$ to $50^\circ$.  Both of these changes result in a small, $\approx$ 1 K increase in the maximum temperature of the nitrogen-ice but in comparison with panel (a), the transport of nitrogen is significantly greater, approximately twice as much mass loss from the pole, for the same time period.  Thus the VT and thermal evolution are sensitive to the thermal parameters.  This sensitivity roots from the sensitive dependence of nitrogen vapor pressure on nitrogen-ice temperature.  Small changes in temperature correspond to larger relative changes in vapor pressure and the higher vapor pressure enables the transport of more material by the atmosphere.  Changes that reduce the maximum temperature of the nitrogen-ice, such as increasing the albedo, significantly reduce the mass transport by the atmosphere.  Since the simulations in figures \ref{fig:ModelOut} and \ref{fig:Mass} were for Eris at aphelion, however, VT will increase dramatically for closer heliocentric distances.  These examples demonstrate that the local atmosphere on Eris could transport significant volatile mass and uniform condensation of a Pluto-like atmosphere may not be able to explain the anomalously high geometric albedo of Eris in the present epoch, several decades past aphelion.  While uniform atmospheric freeze-out may not be able to explain the anomalously high geometric albedo of Eris, seasonal VT remains a plausible hypothesis and will be investigated in a future application of the model.

\begin{figure}[ht!]
  \centering
  \subfloat[Nitrogen-ice Temperature from Coupled Thermal-Transport Model]{\includegraphics[width=0.47\textwidth]{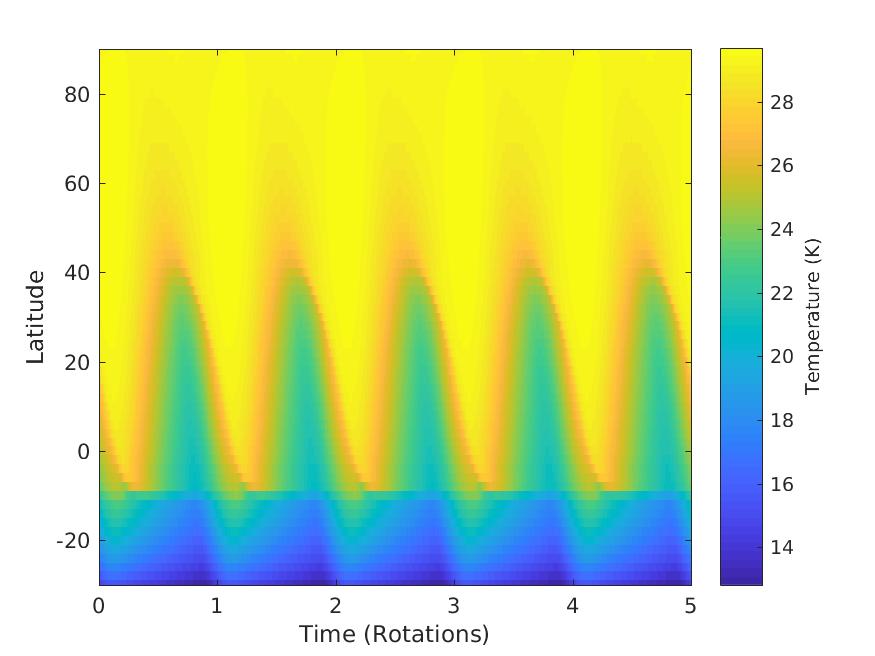}}\hfil                
  \subfloat[Nitrogen-ice Temperature from Thermal-Only Model]{\includegraphics[width=0.47\textwidth]{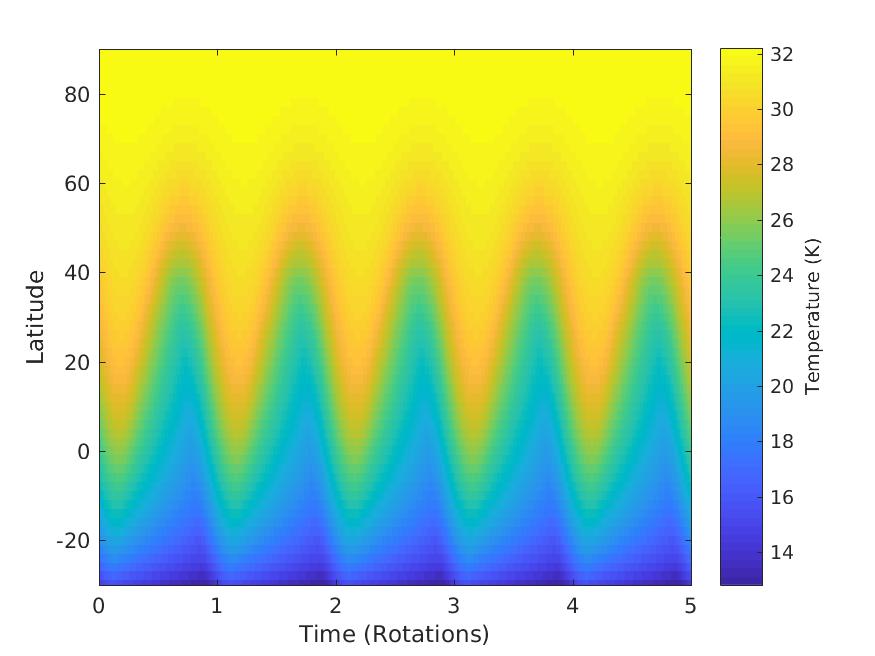}}\hfil
\caption{Comparison of temperature distributions from coupled thermal-transport model (a) and thermal-only model (b).  The initial conditions are the same as in figure~\ref{fig:ModelOut}, except that the bolometric Bond albedo is 0.5 and subsolar latitude is $50^\circ$.  The comparison demonstrates that the transport of volatile material also transports significant energy and tends to homogenize the temperature distribution.  The steep gradient in temperature at approximately $-10^\circ$ latitude in panel (a) marks the location where the vapor pressure becomes too weak to transport appreciable volatile material and the temperature distribution is more similar to the thermal-only distribution.}
\label{fig:TComp}
\end{figure}

Figure~\ref{fig:TComp} compares the temperature distributions from the coupled thermal-transport model and the thermal model without any atmospheric transport.  For the thermal-only case, the transport of volatile material by the atmosphere is ignored by setting the second term of equation~\ref{MassConservation} to zero.  This is an unphysical scenario because the atmosphere will have pressure gradients that will accelerate the vapor and transport material but it is useful for getting a sense of the significance of the transport component of the model.  From comparison of the two temperature distributions, it is concluded that the transport does have a significant effect on the energy balance.  The movement of nitrogen and the latent heat from its sublimation/condensation is effective at transporting energy from warmer to cooler regions and tends to homogenize the volatile-ice surface temperature.  Just as for mass transport, the energy transport depends sensitively on the thermal parameters and increases as temperature increases.  In fact, the modifications from the initial conditions of the simulation in figure~\ref{fig:ModelOut} were selected to increase the apparent difference between the thermal-transport and thermal-only temperature distributions by increasing temperature. 

\section{Discussion}
The first implication to emphasize from the results of the coupled thermal-transport numerical model for a local, collisional, sublimation atmosphere is that volatile transport (VT) on Eris in this regime can be significant.  At a maximum surface pressure of 1-10 nbar, it is tempting to dismiss the atmosphere as negligible with no significant influence on the global energy and mass distributions but the model results demonstrate that this is not the case.  Nitrogen is a very volatile material that sublimates into the vapor phase at the very cold temperatures of the distant Kuiper belt; nitrogen transport, even at heliocentric distances of $\approx 100$ AU, can modify the surface of Eris.  We expect the conclusion that VT is an important surface process will also apply to any other bodies in the Kuiper belt that have substantial nitrogen reservoirs, likely only large bodies \citep{Schaller2007}.  Since carbon monoxide has a similar volatility to nitrogen this statement is also true for any bodies with substantial carbon monoxide reservoirs, although there may not be any bodies in the Kuiper belt where the carbon monoxide inventory exceeds that of nitrogen.

Atmospheric freeze-out (radiative collapse) is a prevalent hypothesis for the anomalously high geometric albedo of Eris (e.g., \citet{Brown2005, Sicardy2011}).  Based on the results shown in figure~\ref{fig:Mass}, however, we conclude that uniform collapse of a global, nitrogen, sublimation atmosphere (that may exist on Eris when it is closer to the Sun, analogous to the atmospheres on Triton and Pluto) likely does not explain Eris's albedo.  This conclusion is true for most bolometric Bond albedos but since the VT depends sensitively on the thermal parameters, it does not hold for extremely high bolometric Bond albedos.  The general hypothesis of seasonal VT (e.g., cycling of volatile-ice between different surface regions each orbital period) remains a plausible explanation for the anomalous albedo of Eris and should be investigated.  The discovery of terrains with exceptionally high albedos on Pluto \citep{Buratti2017} that are renewed primarily by other processes, such as convection and glaciation (e.g., \citet{McKinnon2016, Moore2016}) leads us to propose that geologic processes other than seasonal VT may also be renewing the surface of Eris.  As previously noted, Eris likely has a greater radiogenic heat flux than Pluto, a consideration that increases the plausibility of endogenic resurfacing.

Another implication of the results of the coupled thermal-transport model is that evolution of nitrogen-ice deposits on Eris may result in changes of albedo that are observable with Earth-based telescopes.  The results shown in figure~\ref{fig:Mass} suggest the mass-loss rate from the current summer pole is $\approx$ 1-10 mm of nitrogen-ice per decade.  If the summer hemisphere is covered by nitrogen-ice of a similar thickness and the underlying surface is darker, then the removal of the nitrogen-ice may be detectable in the coming decades.  Note that Pluto's volatile-rich regions are brighter than its volatile-poor regions and are also brighter than Charon's volatile-poor surface \citep{Buratti2017}.  Similarly, VT could result in detectable changes of Eris's color between observations (see \citet{Buratti2011} and \citet{Buratti2015} and references therein for discussions of albedo and color changes of Triton and Pluto from VT). 

Aside from the prediction that the transport of nitrogen-ice may be observable as a change of the albedo or color of Eris, which is based on the results of the model for the volatile-ice mass, the results for the atmospheric pressure may also be testable.  The pressure at the limb can be measured using stellar occultations and we predict a strong variation in pressure with position.  For the parameters given in the caption of figure~\ref{fig:ModelOut}, the maximum pressure above the limb is predicted to be $\approx 1$ nbar.  A previous stellar occultation constrained the pressure at the limb, using two chords, to be $\leq 1$ nbar ($1 \sigma$ confidence level; \citet{Sicardy2011}).  Increasing the chord density and reducing the threshold pressure for detection could lead to the detection of Eris's local atmosphere with a future occultation.

\section{Conclusions}

The $\alpha$ parameter is useful for estimating whether a sublimation atmosphere is in the global regime with approximately uniform surface pressure over the globe or in the local regime with significant horizontal pressure gradients.  Based on this parameter, KBO Eris likely has a global, nitrogen atmosphere at perihelion and a local, nitrogen atmosphere at aphelion.  The $\beta$ parameter (inverse Knudsen number) indicates that the atmosphere on Eris is probably always collisional.

A coupled thermal-transport numerical model developed to simulate thermal and volatile evolution in the local, collisional, sublimation atmosphere (LCSA) regime was introduced.  The model conserves energy, mass, and momentum while maintaining vapor pressure equilibrium.  It is adaptable to any LCSA, an atmospheric regime that occurs on Io and is expected on several objects in the Kuiper belt for parts of their orbits \citep{Young2013}.

The model results indicate that volatile transport (VT) on Eris, even at its aphelion distance of nearly 100 AU, can be significant.  The nitrogen-ice temperatures are $<$ 30 K and vapor pressures are $<$ 10 nbar but the significant pressure gradients in the local atmosphere regime result in transport of nitrogen mass, that integrated over the long timescales associated with such a distant orbit, can be significant as compared to the column mass of the atmospheres of Triton and Pluto.  Although Eris is 96 AU from the Sun in 2018, VT may result in changes in albedo or color that are observable with Earth-based telescopes.  The model predictions for the atmospheric pressure may also be testable using a stellar occultation.

The thermal and volatile evolution in the LCSA regime depends sensitively on the thermal and transport model parameters.  Bolometric Bond albedo, emissivity, rotation period, and rotation pole are important parameters for simulating VT on Eris that are not strongly constrained.  Measurement of these parameters is encouraged.

Uniform collapse of a global sublimation atmosphere, that presumably existed on Eris when it was near perihelion, is probably not the primary reason for its anomalously high geometric albedo in the present epoch, when it is several decades past aphelion.  This conclusion is true for most bolometric Bond albedos but since the VT depends sensitively on the thermal parameters, it does not hold for extremely high bolometric Bond albedos.  The more general hypothesis of seasonal VT remains a plausible explanation for Eris's anomalous albedo.  Other geologic processes such as convection and glaciation that are now thought to be the primary processes renewing Pluto's brightest surfaces (e.g., \citet{McKinnon2016, Moore2016}) are also plausible hypotheses for Eris's high albedo.

\section*{Acknowledgements}
Jason Hofgartner was supported by an appointment to the NASA Postdoctoral Program at the NASA Jet Propulsion Laboratory, California Institute of Technology administered by Universities Space Research Association through a contract with NASA.  This work was inspired by the New Horizons exploration of Pluto.  We thank two anonymous reviewers for their service and helpful comments.

% References
%
% Following citation commands can be used in the body text:
% Usage of \cite is as follows:
%   \cite{key}          ==>>  [#]
%   \cite[chap. 2]{key} ==>>  [#, chap. 2]
%   \citet{key}         ==>>  Author [#]

% References with bibTeX database:

\section*{References}
\bibliographystyle{elsarticle-harv}
\bibliography{Eris_Icarus}

% Authors are advised to submit their bibtex database files. They are
% requested to list a bibtex style file in the manuscript if they do
% not want to use model1-num-names.bst.

\newpage
Copyright 2018.  All rights reserved.

\end{document}